\documentclass[a4paper,fleqn,anonymous]{cas-dc}

\usepackage{graphicx}
\usepackage{subfig} 
\usepackage{float}
\usepackage{amssymb}
\usepackage{algorithmicx}
\usepackage{amsmath}
\usepackage{algorithm}
\usepackage{algpseudocode}

\usepackage{booktabs}
\usepackage{threeparttable}
\usepackage[numbers]{natbib}

\def\tsc#1{\csdef{#1}{\textsc{\lowercase{#1}}\xspace}}
\tsc{WGM}
\tsc{QE}
\tsc{EP}
\tsc{PMS}
\tsc{BEC}
\tsc{DE}

\begin{document}
\let\WriteBookmarks\relax
\def\floatpagepagefraction{1}
\def\textpagefraction{.001}

\title [mode = title]{Detection and Analysis of Sensitive and Illegal Content on the Ethereum Blockchain Using Machine Learning Techniques}   


\author{Xingyu Feng}[type=editor,
                        style=chinese,
                        auid=000,bioid=1,
                        orcid=0000-0001-5340-4592
                        ]

\ead{fxywml@outlook.com}





\affiliation{organization={Hainan University},
    city={Haikou},
    country={China}}

\cormark[1]

\begin{abstract}
Blockchain technology, lauded for its transparent and immutable nature, introduces a novel trust model. However, its decentralized structure raises concerns about potential inclusion of malicious or illegal content. This study focuses on Ethereum, presenting a data identification and restoration algorithm. Successfully recovering 175 common files, 296 images, and 91,206 texts, we employed the FastText algorithm for sentiment analysis, achieving a 0.9 accuracy after parameter tuning. Classification revealed 70,189 neutral, 5,208 positive, and 15,810 negative texts, aiding in identifying sensitive or illicit information. Leveraging the NSFWJS library, we detected seven indecent images with 100\% accuracy. Our findings expose the coexistence of benign and harmful content on the Ethereum blockchain, including personal data, explicit images, divisive language, and racial discrimination. Notably, sensitive information targeted Chinese government officials. Proposing preventative measures, our study offers valuable insights for public comprehension of blockchain technology and regulatory agency guidance. The algorithms employed present innovative solutions to address blockchain data privacy and security concerns.
\end{abstract}


\let\WriteBookmarks\relax
\def\floatpagepagefraction[1]
\def\textpagefraction{.001}
\let\printorcid\relax
\maketitle

\section{Introduction}
\label{1}
In recent years, blockchain technology has attracted widespread attention from academia and industry and is hailed as a new technology that will trigger social changes. Simply put, blockchain technology is a new type of distributed ledger technology that can realize trusted transactions without intermediaries in an environment of mutual distrust. Different from the traditional information system, the blockchain system has many characteristics such as user anonymity, transaction traceability, anti-counterfeiting, non-tampering, and the ability to realize smart contracts. It has been widely used in health care~\cite{1}, financial contracts~\cite{2}, enterprise operation management, data management, auditing~\cite{4}, and many other fields, showing an explosive growth trend.

Due to the broad application prospects of blockchain technology, academia has carried out various research on this technology, such as consensus mechanism, privacy protection~\cite{8}, system security~\cite{10}, and smart contract security~\cite{13}, etc. Since the blockchain system cannot tamper with the user's various behavior records and operation results, it has become an important research direction to analyze user behavior and discover potential social risks based on blockchain data~\cite{7}. Since the blockchain system has the characteristics of anonymity and immutability and is usually associated with digital assets (such as various coins, tokens, etc.), there may be various abnormal and illegal behaviors on the blockchain platform. Therefore, establishing illegal behavior identification and early warning models based on blockchain data has become an important topic in the field of blockchain data analysis. At present, researchers have established a variety of illegal behavior identification models, such as phishing fraud identification~\cite{14}, abnormal account analysis, Ponzi scheme identification~\cite{15,35}, etc. However, these illegal acts exist essentially because of the existence of "coins". On some industry-oriented blockchain platforms, coins do not necessarily exist, but the essence of blockchain as an information system has not changed. Therefore, rather than identifying various anomalies and illegal behaviors for "coins", a more general question is: Is there any abnormal or illegal information in the blockchain system?

The blockchain mainly realizes the reliable accounting of digital events, such as the application of financial applications and asset certification services via the blockchain~\cite{16}. But at the same time, due to the anonymous, open and immutable nature of the blockchain, anyone can embed all kinds of information in Ethereum transactions. In earlier studies, researchers found that child abuse, pornography~\cite{17}, etc. existed in Bitcoin transactions, which seriously affected the reputation of the Bitcoin system. Due to the limitations of Bitcoin transaction fields, a large amount of information cannot be embedded in a Bitcoin transaction at one time. In the Ethereum blockchain, the amount of information a transaction can embed is greatly increased. A natural question is: Is there any behavior of illegal, sensitive, malicious information embedding in Ethereum transactions? What specific harmful information is embedded and what impact it might have? The answers to these questions will help us to further understand the application status of the blockchain system, and then strengthen the supervision measures of the system during its application.

\begin{figure*}[!t]
  \centering
  \includegraphics[width=6in]{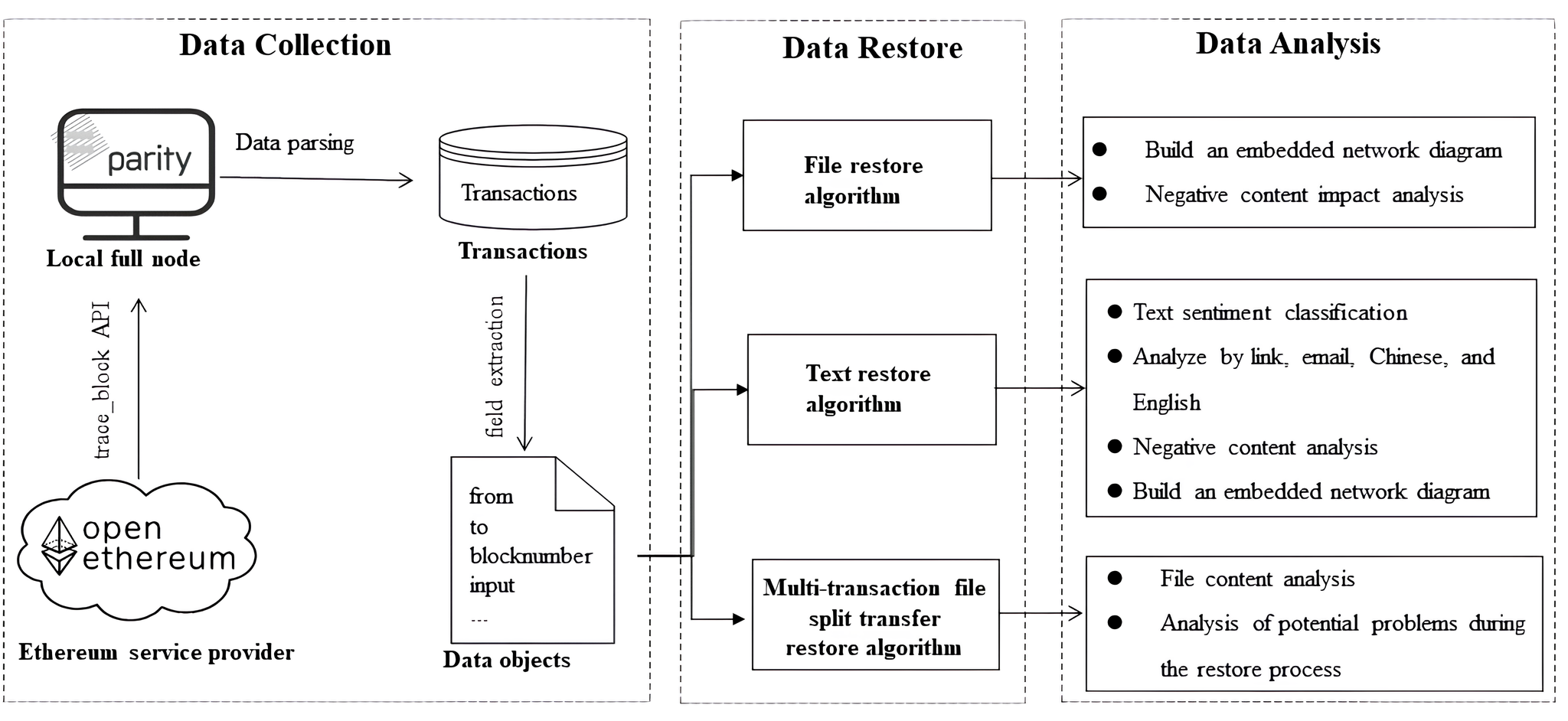}
  \caption{A framework for embedding restoration and analysis of data in the Ethereum blockchain network.}
  \label{fig1}
\end{figure*}

At present, blockchain technology has been incorporated into the construction of new infrastructures in countries around the world. In the future, blockchain technology will become an important supporting technology for the global digital economy. Strengthening the supervision of blockchain platforms and technologies is an inevitable measure for blockchain applications. In this context, this paper conducts a comprehensive analysis of the possible malicious information embedding behavior on the Ethereum platform for the first time. Specifically, as shown in Fig.\ref{fig1}, the original data in the chain was first obtained through the Ethereum client Parity, and the corresponding fields of the transaction data are extracted; On this base, we developed algorithms to restore the files and text data and algorithms to reverse the file splitting and embedding in multiple transactions; We analyzed about 3.4 billion transaction records on Ethereum, and restored multiple files, texts, pictures, and other related data; Then, the Fasttext algorithm in natural language processing is used to analyze the sentiment of the embedded text data, and the open-source NSFWJS library is used to identify pornographic images. Finally, we carried out an in-depth analysis of the restored data from various aspects and summarized the possible impact of these various types of information.

This study aims to identify, restore, and analyze data embedding on the Ethereum platform by designing an algorithm for data identification and analyzing approximately 300 million transaction records. We utilized the FastText algorithm to classify text data sentiment and identify potentially sensitive or illegal information, and the NSFWJS library to detect indecent images embedded in the transactions. Our findings reveal the existence of harmful content on the Ethereum blockchain and we propose measures to prevent the spread of such data in the future(refer to Section 6). This study is the first comprehensive exploration and analysis of embedded content in the Ethereum blockchain and offers valuable insights for the public and regulatory agencies. Additionally, the algorithms and technologies employed may provide new methods for addressing data privacy and security concerns on the blockchain.

Compared with existing blockchain-based data analysis work, the main contributions of this paper are: 
\begin{itemize}
\item {\verb||}In this study, for the first time, the data embedded in the Ethereum blockchain was analyzed and restored comprehensively (as of the 13 millionth block), and a total of 175 common types of files, 296 image files, and 91,206 pieces of text data were found.
\item {\verb||} The study proposed recognition and restoration algorithms for ordinary text, file data, and image data in blockchain transactions based on feature codes and text-encoding methods.
\item {\verb||} The research utilized natural language processing algorithms to classify the text data embedded in the blockchain and identify pornographic images. The study discovered various embedded sensitive and illegal information, pointing out the potential risks of the Ethereum network and providing robust evidence for the necessity of blockchain supervision.
\end{itemize}

The structure of this paper is organized as follows: In Section\ref{2}, we provide technical background on the Ethereum blockchain and its characteristics. Section\ref{dataCollection} discusses the data collection work for this paper. Section\ref{4} presents the design and implementation of three algorithms that we propose for identifying and restoring embedded data in the Ethereum blockchain.  In Section\ref{5}, we analyze the content of the restored data, explore the types of data that are commonly embedded, and relate our findings to the research objectives outlined in Section\ref{1}.  Section\ref{6} discusses the risks and prevention measures of embedding harmful data, including the potential impact on user privacy and system security. Section\ref{3} reviews related work in this area, highlighting the strengths and limitations of existing studies and analyzing their relationship with our work. Finally, in Section\ref{7}, we conclude the paper by summarizing our key findings and contributions and providing specific suggestions for future research in this area.

\section{Background}
\label{2}
\subsection{Ethereum and smart contracts}
Ethereum is the largest programmable blockchain platform in the blockchain world. At the end of 2013, Vitalik Buterin~\cite{brody2021ideologies}, the founder of Ethereum, released the Ethereum white paper, which introduced smart contracts into the blockchain, opened the application of blockchain outside the field of encrypted digital currency, and marked the beginning of the blockchain 2.0 era~\cite{sood2022revamping}. By introducing the Ethereum Virtual Machine (EVM), it can support smart contracts of any complexity. To incentivize peers to participate in the maintenance of the system and to prevent potential abuse of the system, a cryptocurrency called ether (ETH) was created. It is now the second-largest cryptocurrency by market capitalization.  On the blockchain, a smart contract can be viewed as a type of rule-based, automatically executable computer code, or as a digital version of a traditional contract. The introduction of smart contracts has made blockchain technology show more possibilities, and Ethereum has also become an important blockchain platform due to its commitment to building the underlying platform for smart contracts~\cite{taherdoost2023smart}.

\subsection{Ethereum blockchain transaction fields }

In Ethereum blockchain transaction records, the "value" field represents the amount of ether transferred. The "Gas limit" field indicates the maximum amount of Gas (minimum 21,000) that the user is willing to pay to perform an operation or confirm a transaction~\cite{laurent2022transaction}. The "Gas price" field indicates the price the user is willing to spend per gas unit. When making each transaction, the sender sets the Gas Limit and Gas Price, and the Gas Limit*Gas Price gets the cost of the ETH transaction commission. In particular, part of this cost depends on the amount of information contained in the data fields~\cite{33}.

A detailed description of key fields that can store information in Ethereum transaction records is shown in Table \ref{tab:txFields}. In the transaction record, the "input" field can embed up to 700KB of data, while other fields can only embed up to 32KB of content. Therefore, only the "input" field in the transaction record is considered in this paper. The content stored in this field is a hexadecimal string that can represent any type of data. When sending transactions to the Ethereum network, data such as images, videos, and files can be converted into hexadecimal and embedded in the "input" field as incidental information and sent with the transaction. Once the transaction records are successfully uploaded to the chain, these data are permanently stored on the Ethereum blockchain. Currently, most websites like Etherscan.io can display the text information embedded in the "input" field in the transaction record so that this information can be propagated on the blockchain.

\begin{table*}[!t]
\centering
  \caption{The key transaction fields storage size description of Ethereum.}
  \label{tab:txFields}
  \begin{tabular}{cccc}
    \toprule
    Type & Name & Description & Length \\
    \midrule
    Address & From & The sender account & Up to 20 bytes \\
    Address & To & The recipient account, if empty, the transaction will create a contract & Up to 20 bytes \\
    Quantity & Gas limit & The amount of gas provided to execute this transaction & Up to 32 bytes \\
    Quantity & Gas price & Integer of the gas price used for each paid gas & Up to 32 bytes \\
    Quantity & Value & Integer of the value sent with this transaction & Up to 32 bytes \\
    Data & Input & The data sent along with the transaction & 0 - about 700KB \\
    \bottomrule
  \end{tabular}
\end{table*}

\section{ Data collection}
\label{dataCollection}
In order to collect the relevant data of the input field, this article utilized the Parity client officially provided by Openethereum to construct a local full node of the Ethereum network for synchronizing the transaction data of the entire network. Open Ethereum provides numerous JSON-RPC APIs for locally synchronized data acquisition. By employing the official trace\_block API, we can access information regarding all transactions in a block based on the block number.

Using this interface, we gathered data from 13 million Ethereum blocks, covering the period from November 1, 2015 (the launch date of the Ethereum main net) to August 10, 2021, using the provided interface. The dataset comprises around 3.4 billion transaction records. Subsequently, we conducted an analysis on the dataset, focusing on extracting the content from the input field of each transaction record.

To analyze the data embedded in the "input" field, it is crucial to convert it into human-readable information. Since Ethereum places no restrictions on the type of data that can be embedded in this field, we believe that there may be three types of content: (1) files, such as image files in JPG, PNG format, PDF files, etc.; (2) plain text content, such as Chinese, English, email, links, etc.; (3) "meaningless" content, such as characters input randomly by the transaction initiators, bytecodes compiled by the contract, etc. Therefore, it is necessary to design algorithms that can recognize the various possible situations to analyze the embedded files and text data, ultimately achieving the "restoration" of the embedded data. Additionally, since a piece of data may be divided into multiple transactions for embedding due to the limitations of Ethereum gas price and field capacity, the data restoration algorithms must also takeconsider this split-embedding scenario the next section, we will focus on the related algorithms designed.

\section{Data restoration algorithm}
\label{4}
This article uses the UTF\-8 encoding method in the Unicode encoding rules to decode text data.
When the data embedded in Ethereum does not use the sub-encoding method, there will be some "gibberish characters" during the decoding process, that is, characters that are not included in the UTF\-8 encoding. To filter out these "gibberish characters", according to the principle of UTF\-8 encoding, "gibberish characters" filtering rules are designed: if the continuous bytes satisfy the UTF-8 encoding, they are kept, otherwise, they are directly discarded. The specific process is shown in Algorithm\ref{alg:utf8-filter}. In this paper, for the sake of familiarity, this paper only considers decoding Chinese and English texts. The specific text restoration algorithm is shown in Algorithm\ref{alg:text-restore}. We compared the work on blockchain-embedded data recovery in Table \ref{tab:Comporison_Restoration}, and our recovery methods extract more and richer content.

\begin{table*}[!t]
\centering
  \caption{Comparison of related work on Blockchain-embedded data restoration.}
  \label{tab:Comporison_Restoration}
  \begin{tabular}{|c|c|c|c|c|}
    \hline
    Related Work & \parbox{2cm}{Insert methods} & \parbox{2cm}{Restore methods} & \parbox{3cm}{Dataset size} & \parbox{3cm}{Restore Category}\\
    \hline
    Roman Matzutt~\cite{17} & \parbox{2cm}{Bitcoin Blockchain (coinbase and OP RETURN)} & \parbox{3cm}{(1) ASCII characters\\ (2)Service Detectors} & \parbox{4cm}{Bitcoin blockchain (On August 31st, 2017, which was containing 482,870 blocks and 200 million transactions)} & \parbox{3cm}{1557 files (59 files, 92.25\% Text)}\\
    \hline

    Teppei Sato~\cite{31} & \parbox{2cm}{Ethereum Blockchain (the “inputdata” field)} & \parbox{3cm}{Foremost tool} & \parbox{4cm}{Ethereum network (from 0 to 6,988,614 in block height) (July 30, 2015–December 31, 2018 UTC)} & \parbox{3cm}{154 files (jpg, png, gif, htm, zip, pdf, and exe)}\\
    \hline

    Ours & \parbox{2cm}{Ethereum Blockchain (the “inputdata” field)} & \parbox{3cm}{(1) File restore algorithm\\ (2) Text restore algorithm\\ (3) Multi-transaction file split transfer restore algorithm} & \parbox{4cm}{Collected data from 13 million Ethereum blocks, spanning from November 1, 2015 (the Ethereum main net launch date) to August 10, 2021, about 250 million transactions} & \parbox{3cm}{(1) 175 common files (jpg, html, zip, pdf)\\ (2) 296 image files\\ (3) 91,206 text files.}\\
    \hline
  \end{tabular}
\end{table*}

\begin{algorithm}
    \caption{Text Restore Algorithm}
    \label{alg:text-restore}
    
    \textbf{Input}: Hex string $input\_hex$ 
    
    \textbf{Output}: Restored text $restore\_text$
    
    \begin{algorithmic}[1]
        \State Define UTF-8 Chinese coding range $utf8\_range\_c$
        \State Define English coding range $utf8\_range\_e$
        
        \State Convert $input\_hex$ to byte array $hex\_byte\_array$
        \State $filtered\_bytes$ = UTF8Filter($hex\_byte\_array$) \label{line:call-utf8-filter}
        
        \If{$filtered\_bytes$ contains characters in $utf8\_range\_c$} 
            \State $restore\_text =$ RevertToChineseText($filtered\_bytes$)
        \ElsIf{$filtered\_bytes$ contains characters in $utf8\_range\_e$}
            \State $restore\_text =$ RevertToEnglishText($filtered\_bytes$)
        \EndIf   
        
        \Return $restore\_text$
    \end{algorithmic}  
\end{algorithm}


        





\subsection{File restoration algorithm}
There are various types of files on the computer, including jpg, png, docx, etc. Usually, the type of a file is judged by the suffix of the file, that is, the extension of the file. However, the extension of the file can be modified artificially. If the suffix of the jpg file is changed to docx, the file will not be opened. After changing the file suffix, a new method is required to restore its original file type. Specifically, the file type can be determined by monitoring the hexadecimal format of the file. In the header of the hexadecimal data of the file, there is a hexadecimal code called the file signature. This hexadecimal code is usually a fixed number of bytes, and different files have different feature codes. Usually, one only needs to look at the contents of these few bytes to know the type of the file. With these file signatures, one can directly determine the type of a file without knowing the file extension.

\begin{table}[!t]
\centering
  \caption{Common file signatures}
  \label{tab:file_sig}
  \begin{tabular}{cc}
    \toprule
    Type & File Signatures \\
    \midrule
    png & 89 50 4E 47 0D 0A 1A 0A \\
    jpg & FF D8 FF E0 \ FF D8 FF E1 \ FF D8 FF E8 \\
    html & 68 74 6D 6C 3E \\
    zip & 0 4B 03 04 \ 50 4B 4C 49 54 45 \ 57 69 6E 5A 69 70 \\
    pdf & 25 50 44 46 \\
    \bottomrule
  \end{tabular}
\end{table}

Table\ref{tab:file_sig} lists common file signatures. For example, the file feature codes in jpg format are "ffd8ffe0", "ffd8ffe1", and "ffd8ffe8", and the file feature code in png format is "89504e47". In addition, some image files also have signatures at the end, for example, the signature at the end of a gif file is 0x003B. Based on this, it can be judged whether the end of the file has been reached. In the subsequent file segmentation and embedding restoration algorithm, the end feature code of the image file will be used as the file end mark.

To identify files that are more likely to exist in the data, first, we use regular expression matching to detect the data containing the file signature, and then discard the part before the signature (if there is a file type with a trailing signature, one need to discard the part after the trailing signature), and then use the Java stream tool to convert the data into a byte stream, and finally restore the byte stream to the corresponding file according to the file type represented by the feature code. This completes the preliminary restoration of the file data. The detailed implementation is shown in Algorithm\ref{alg:file-restore}.

\begin{algorithm}
    \caption{File Restore Algorithm}
    \label{alg:file-restore}
    
    \textbf{Input}: Hexadecimal string $input\_hex$ 
    
    \textbf{Output}: Restored text data $restore\_file$
    
    \begin{algorithmic}[1]
        \State Load file header signature $file\_head\_sig$ 
        \State Load file trailer signature $file\_tail\_sig$ 
        
        \If{$file\_head\_sig$ is found in $input\_hex$} 
            \If{$file\_head\_sig \& file\_tail\_sig$ are both found in $input\_hex$}
                \State $input\_hex =$ Hex between the two signatures
            \Else
                \State $input\_hex =$ Hex after $file\_head\_sig$
            \EndIf
        \EndIf 

        \State Convert $input\_hex$ to byte array $input\_byte\_array$
        \State $restore\_file =$ Restore file from $input\_byte\_array$
        
        \Return $restore\_file$
    \end{algorithmic}  
\end{algorithm}

\subsection{Multi-transaction-based file segmentation and embedding restoration algorithm}
Considering that a file may be divided into multiple transactions and embedded separately because of its size, this paper designs a segmentation and embedding restoration algorithm. That is, once a transaction that embeds incomplete files is recognized, multiple transactions sent by the same sender after the first transaction are searched. Algorithm\ref{alg:file-restore} is called to restore possible files by synthesizing data from multiple transactions. In this article, considering that Ethereum generates a block in about 13 seconds, a relatively complete file is restored by looking for 6700 blocks later, that is, all transactions sent by the sender within about 24 hours after sending the initial transaction. The detailed implementation is shown in Algorithm\ref{alg:file-Split-restore}.

\begin{algorithm}
    \caption{File Split Embedded Restore Algorithm}
    \label{alg:file-Split-restore}
    
    \textbf{Input}: Incomplete file input data $input\_hex$; Current $block\_number$ 
    
    \textbf{Output}: Restored file $seg\_restore\_file$
    
    \begin{algorithmic}[1]
        \State Load file header signature $file\_header\_sig$ 
        \State Load file trailer signature $file\_trailer\_sig$ 
        
        \If{$file\_header\_sig$ is found in $input\_hex$} 
            \For{$i = block\_number; i \geq 0; i--$}
                \If{$file\_trailer\_sig$ is found in $input\_hex$}
                    \State Extract the substring between the head and tail signatures
                    \State Call file restore Algorithm\ref{alg:file-restore}
                    
                    \Return $seg\_restore\_file$
                \Else
                    \State $input\_hex \mathrel{+}= input\_hex$
                \EndIf
            \EndFor
            \State Call file restore Algorithm\ref{alg:file-restore}
            
            \Return $seg\_restore\_file$
        \EndIf 
    \end{algorithmic}  
\end{algorithm}

\section{Data embedding analysis}
\label{5}
Using the aforementioned data acquisition methods and data restoration algorithms, about 3.4 billion transaction data were analyzed, and various forms of data were decoded, including 175 common types of files and 296 images. Table\ref{tab:restore_result} summarizes the types and quantities of text data. Among them are 73,180 English texts; 17,053 Chinese texts; 944 email addresses and 831 links. The restored data will be analyzed in detail in the following part.
\begin{table}[!t]
\small
\setlength{\tabcolsep}{0.08cm}
  \caption{Restoration result statistics}
  \label{tab:restore_result}
  \begin{tabular}{ccccccc}
    \toprule
    Com files & Img  & English& Chinese & Mail & Link \\
    \midrule
    175 & 296 & 74163 & 17043 & 944 & 831 \\
    \bottomrule
  \end{tabular}
\end{table}

\subsection{File data analysis}
\begin{figure}[!t]
  \centering
  \includegraphics[width=3in]{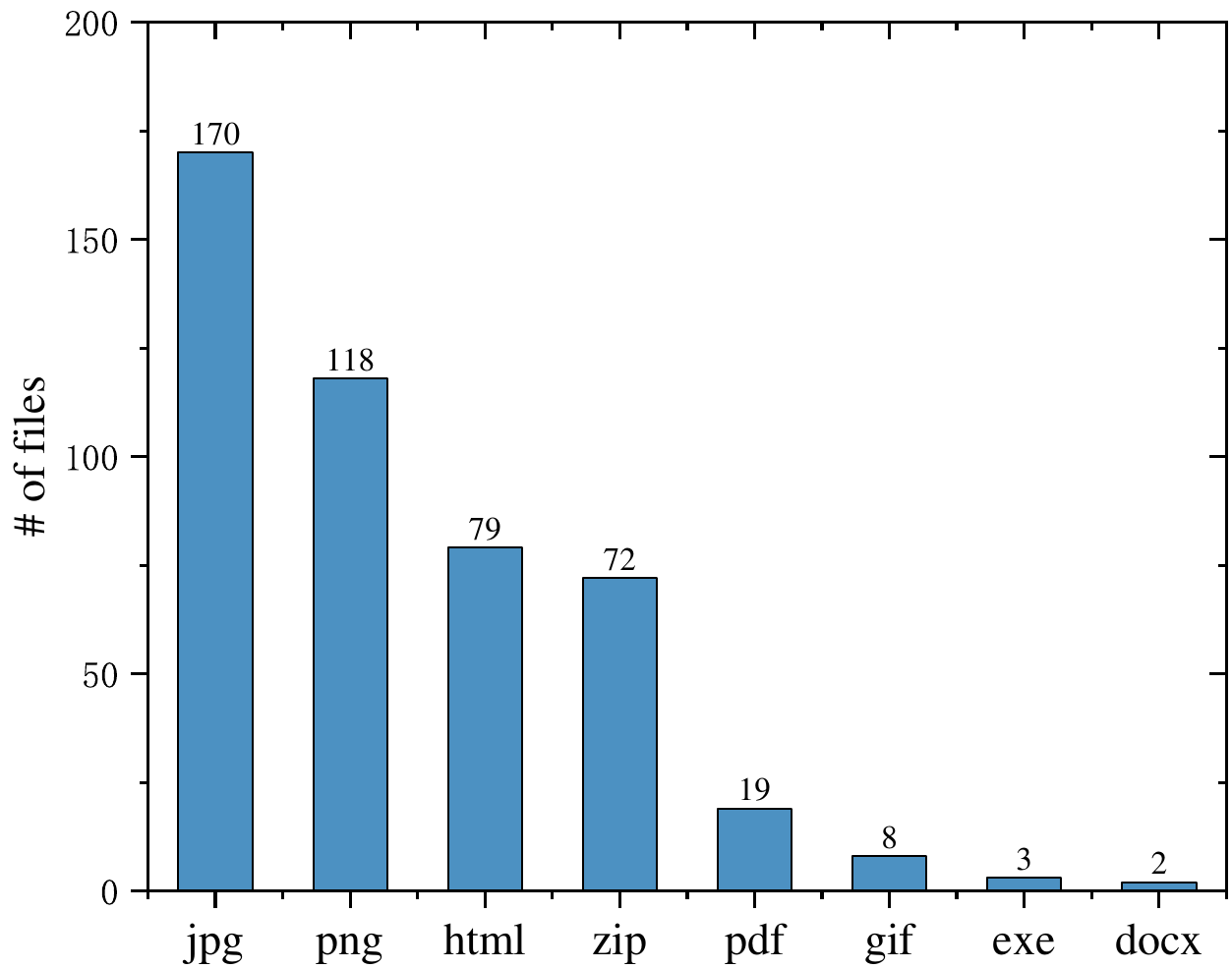}
  \caption{The file types and quantities are restored}
  \label{fig2}
\end{figure}

\begin{figure}[!t]
  \centering
  \includegraphics[width=3in]{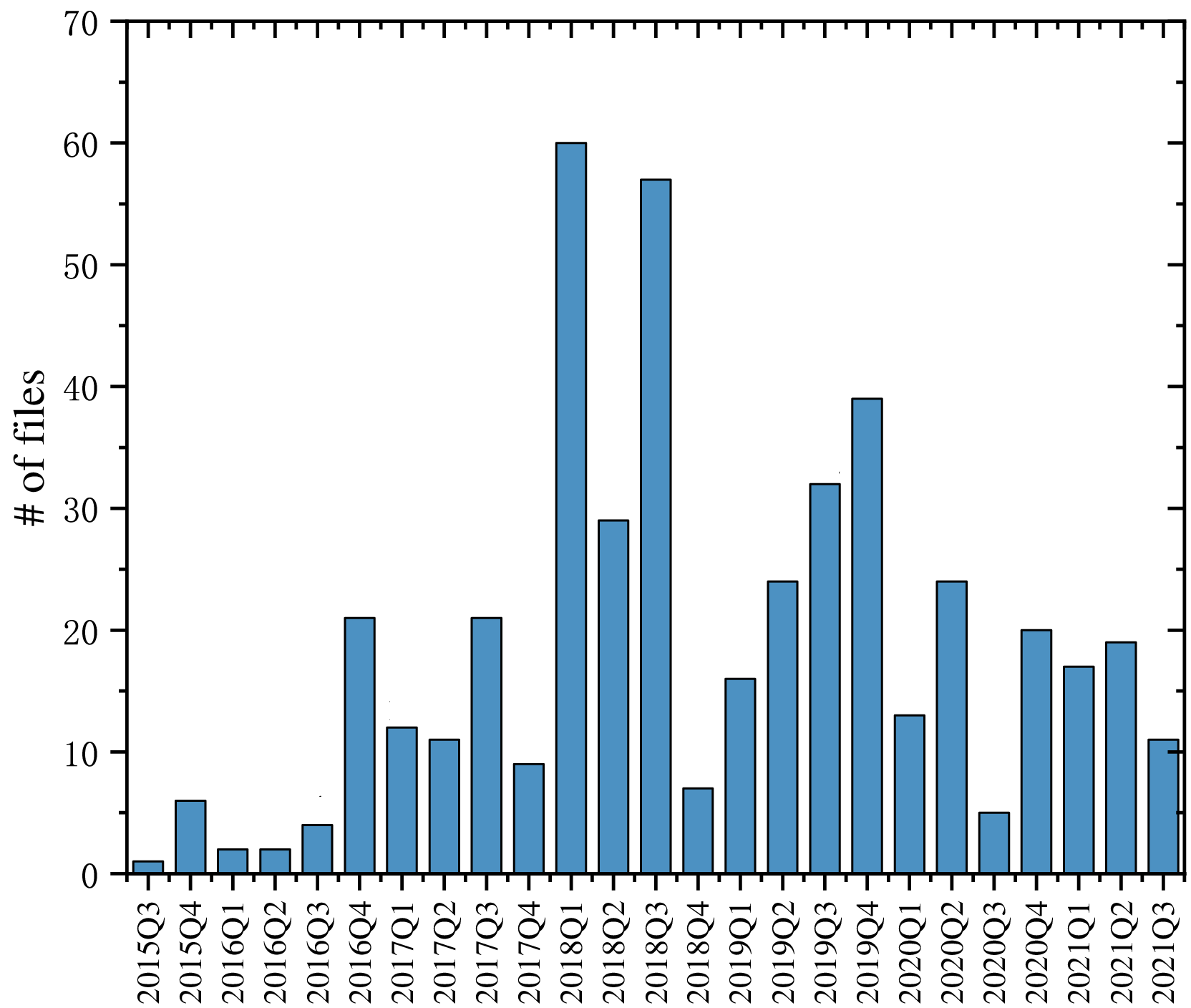}
  \caption{Number of embedded files per quarter (2015-2021).}
  \label{peer_mouth}
\end{figure}

This section detects 19 common file types. Specifically, they include jpg, gif, png, zip, exe, xls, docx, html, css, js, flv, mp4, mp3, wav, jsp, xml, sql, java, properties, and other file types. Figure \ref{fig2} shows the types and numbers of files embedded in the Ethereum nich obviously shows that image is the most commonly embedded file format. Additionally, an examination of the number of embedded files over the quarters (2015-2021) was conducted, as depicted in Figure \ref{peer_mouth}. It was observed that a significant number of files were incorporated into the Ethereum network through transactions during the years 2018 and 2019. The following is a detailed analysis of the restored files containing undesirable content.

\subsubsection{Image file data analysis}
By identifying and restoring the files embedded in the transaction, 175 common types of files and 296 images were finally obtained. As shown in Fig.\ref{fig2}, 63\% of the embedded files belong to images. Among them, 231 are complete images and 65 are incomplete ones. To further analyze the information of the images, all the images that have been received more than 4 times by the receiving addresses were further analyzed. The specific results are as follows:

(1) Thirty pictures are NFT-encrypted artwork sent to the CreepyCryptos$\footnote{https://twitter.com/creepy\_cryptos}$ contract address; 55 are personal records, such as group photos of people, selfies, and funny images. Although these pictures are normal, they leak the user's privacy on the chain. Analyzing these pictures makes it easy to obtain the holder information of the relevant Ethereum address. The remaining images include park icons, cartoon characters, natural scenery, etc.

(2) We call the open-source NSFWJS library$\footnote{https://nsfwjs.com/}$ to identify pornographic images. NSFWJS is a simple JavaScript library that uses TensorFlow.js to perform content detection locally on the client side and quickly identify indecent images. The project is open source. Currently, the recognition accuracy of the open-source library is about 90\% in the small model and about 93\% in the medium model. Therefore, this experiment uses this library for pornographic image detection. There are seven erotic images. For ethical reasons, citations are avoided here. We consider embedded pornographic images objectionable in almost all jurisdictions. The pictures contain four female nudes and three male nudes, and one contains an illegal URL link.

(3) It is also noteworthy that there is a Nazi flag emblem composed of black, white and red colors in the picture file. The icon is banned from public places in most European countries. Now the Nazi flag emblem is embedded in the Ethereum network for permanent preservation, which increases the challenges for prohibiting the spread of Nazi ideas.

\vspace{0.2cm}
\framebox{%
\begin{minipage}{0.9\linewidth}
\textbf{Finding 1:}Most of the images embedded in the Ethereum network are recording normal content. These may lead to the disclosure of the privacy of users in the chain. It is noteworthy that there are also harmful images on the chain that cannot be ignored, involving pornography, terrorism, etc.
\end{minipage}}
\vspace{0.2cm}

\subsubsection{Other types of file analysis}
Since other types of files other than image files account for 37\% of the total number of files, they will be collectively referred to as other types of files in this section. Including 72 compressed files, 79 html files, 19 pdf files, 3 exe files, and 2 docx files. The detailed analysis is as follows:

(1) Among the 72 compressed files restored, 30 encrypted files were found. How to decrypt the content is beyond the scope of this article and our profession, so we cannot know the specific content of the file, but it is unusual for a large number of encrypted files to be embedded in a blockchain platform. An easy guess is to use the Ethereum network to engage in some planning for illegal activities. Among the remaining 42 openable compressed files, 40 contain Russian content, and they were embedded in the Ethereum network in April 2019. The content includes php files, html files, personal blogs, literary works, and related speeches involving religion.

(2) The restored 3 exe files were embedded in the blockchain by the same account, which was the same as the finding in~\cite{31}. The files were detected using the VirusTotal website, which is a website that provides a free suspicious file analysis service. In the end, it was found that two exe files were the same and their probability of being a virus file was 80\%, and the probability of the other file being a virus file was 87\%.

\vspace{0.2cm}
\framebox{%
\begin{minipage}{0.9\linewidth}
\textbf{Finding 2: }In addition to pictures, various other types of file embedding in the Ethereum network may involve virus transmission, privacy leakage, copyright infringement, etc. An unexpected finding is the presence of the embedding of encrypted files. One possible reason for their existence is the planning for illegal activities. This could also be used to timestamp some essential documents or bypass some countries' censorship.
\end{minipage}}

\subsection{File segmentation embedding analysis}
\begin{figure}[!t]
  \centering
  \subfloat[Incomplete image files]{
  \includegraphics[scale=.5]{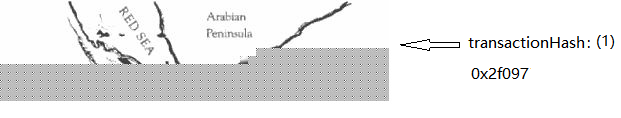}
  \label{fig4(a)}
  }
  
 \subfloat[Algorithm\ref{alg:file-Split-restore} restores the image files]{
  \includegraphics[scale=.5]{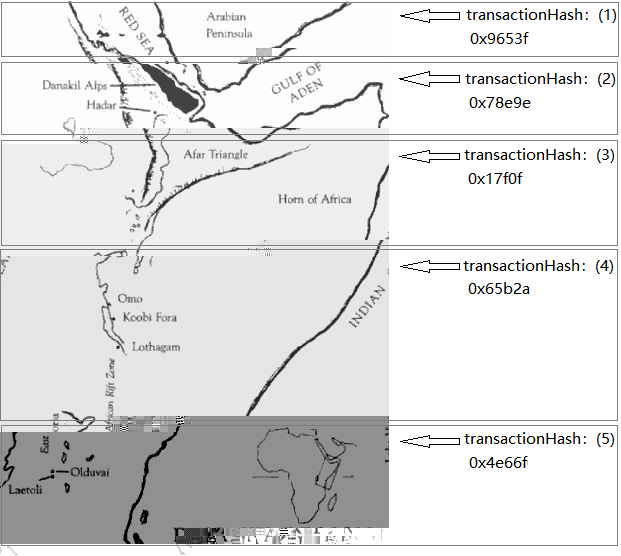}
  \label{fig4(b)}
 }
 \caption{Incomplete image restoration using file segmentation embedded restore algorithm (Algorithm\ref{alg:file-Split-restore}), (a) incomplete image file found, and (b) image file restored using Algorithm\ref{alg:file-Split-restore}. The image file is split 5 times for embedding, and the top 5 bits of the hash value of each transaction are marked in the figure.}
 \label{fig4}
\end{figure}

Due to transaction field capacity limitation, a large file may be embedded in the blockchain network separately through multiple transactions. To this end, file segmentation and embedding restoration algorithm (Algorithm\ref{alg:file-Split-restore}) is designed. Fig.\ref{fig4} shows an example of applying algorithm\ref{alg:file-Split-restore} to restore a segmented and embedded image. This image is a map of East Africa. The incomplete image was identified using the header signature of the file (Fig.\ref{fig4(a)}). After the header signature, the image header file was included, which stored the image size, pixel number, image resolution, etc. According to the information contained in the header file, even if the image is segmented and embedded, the original size of the image can still be restored in the end (to avoid occupying a large space, Fig.\ref{fig4(a)} only shows the partial deletion). In Fig.\ref{fig4}, the image is completely stored in the Ethereum network after five splits, and the top five digits of each transaction hash value are marked on the right side.

In the restored Fig.\ref{fig4(b)}, it can be found that the restored pictures cannot be completely connected at the splits, and each split has different degrees of color changes. To find out the underlying reasons, we checked the complete transaction information on the Etherscan website and found that the receiving address of the transaction was a smart contract with a data upload function. Since the $upload Data(bytes\_data)$ method in the contract needs to be called each time to upload data, the called method name will also be converted into a hexadecimal string and written into the "input" field. Since Algorithm\ref{alg:file-Split-restore} is based on the header and tail feature codes of the file to restore it, other data in the middle of the file cannot be identified, which finally leads to the phenomenon of incomplete connection at the split. For the color change after splitting, the reason has not yet been identified. We speculate that it may be related to calling the smart contract. Probably the input of strings after calling functions causes this phenomenon because it was never found in the split embedded pictures that do not call the smart contract. The restoration example in Fig.\ref{fig4} proves our speculation about multi-transaction file embedding, and the multi-transaction-based file segmentation and embedding restoration algorithm designed in this paper can effectively restore segmented and embedded files.

After further analysis, we found 54 incomplete pictures, among which 10 pictures were segmented and embedded. After applying the file segmentation and embedding restoration algorithm, six complete pictures were obtained, including 1 blockchain proof of work map, 1 map of East Africa, 1 picture of human eyes, 2 landscape pictures, and 1 cartoon bird. At the same time, in the pdf file that cannot be opened, three files are successfully restored, which are the Ethereum white paper, a research paper about gorillas, and the Foxit software advertisement.

\vspace{0.2cm}
\framebox{%
\begin{minipage}{0.9\linewidth}
\textbf{Finding 3: }In terms of information embedding, splitting and embedding large files or long information is a feasible means of on-chain information embedding. But this increases the difficulty of file restoration for monitoring and analysis
\end{minipage}}

\subsection{Text data analysis}
\begin{figure}[!t]
  \centering
  \includegraphics[width=3in]{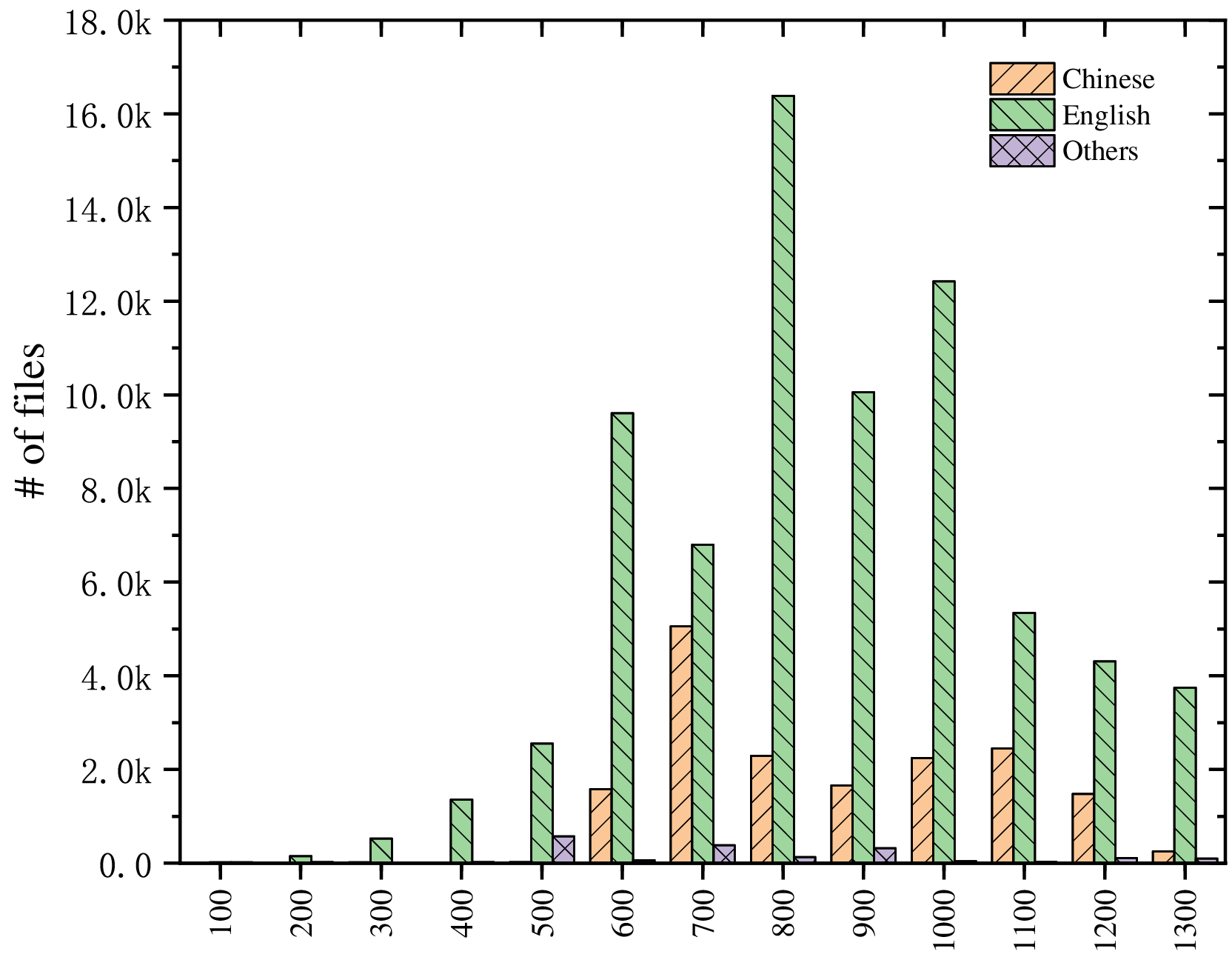}
  \caption{Number of embedded texts per 1 million blocks in the Ethernet blockchain.}
  \label{peer_100w}
\end{figure}

In this section, the directly embedded text data are classified and we focused on four types for analysis: Chinese, English, URL, and email. In order to scrutinize the temporal features embedded within the textual data, the number of text embeddings into the Ethereum network was categorized based on Chinese, English, and other types, and subsequently tallied per every one million blocks. As illustrated in Figure \ref{peer_100w}, a notable surge in text embeddings can be discerned within the range of 5 to 10 million blocks (spanning the years 2017 to 2021). In terms of embedding types, it is evident that English texts undeniably constitute the predominant portion, yet Chinese texts also account for a substantial presence. Subsequently, a more in-depth analysis of the content and characteristics of these various text categories will be conducted.

\subsubsection{Link type text analysis}
There are 720 email addresses sent by these senders to participate in the ICO. 96 among them have "http" in the URL and 218 among them have the identical URL "podcrypt. app". 93 among them have the identical URLs "ddblock.io", which is the official website of slideshow software. It is noteworthy that there is also a URL ending with ".onion". This is the onion browser URL format for accessing the dark web. We did not open it for further analysis, but the dark web URL itself suggests that the link may contain undesirable content.
\subsubsection{Text sentiment classification}
In 2016, Facebook proposed the FastText~\cite{34} text classification model for the first time. Its core idea is to superimpose and average the words of the entire document and the N-gram model vectors to obtain document vectors and then use the document vectors for multi-classification. Compared with the deep neural network method, the FastText model has the characteristics of small network parameters, fast training of the model, and high classification accuracy. In this experiment, we first select the existing emotion-labeled data set published on Chinese Weibo social media$\footnote{https://github.com/BT5153-Group-Seventeen/Weibo-Sentiment-Analysis-During-COVID-19}$(the Chinese social media similar to Twitter) and the English Twitter emotion-labeled data$\footnote{https://www.kaggle.com/datasets/gpreda/all-covid19-vaccines-tweets}$ set to train the FastText model and then optimize the model's parameters to improve the classification accuracy. Then The text data restored in this paper is classified into emotions, and finally, the data of different emotion categories are analyzed.

We use the public Weibo dataset with emotional labels and the Twitter sentiment classification dataset to train the model. There are three data labels: neutral, positive, and negative. Using the above public data to train the FastText model, the trained model has a classification accuracy of 0.925 for Chinese and 0.949 for English data. Then the trained FastText model is used to classify our data, which contains 17,050 Chinese texts and 74,163 English texts restored in this article. The experimental results are shown in the table\ref{tab:sentiment_data}. The Chinese text contains 3,479 negative emotions, 4,155 positive emotions, and 9,450 neutral emotions. The English text has 12,331 negative emotions, 1,093 positive emotions, and 60,739 neutral emotions. In the table\ref{tab:sentiment_display}, we randomly sampled the classification results of texts with different emotional labels and made a display. The last column shows the probability that the text belongs to the label, and the probability threshold can be adjusted freely according to the need to realize personalized screening.

\begin{table}[!t]
\centering
  \caption{Text Sentiment Classification Results}
  \label{tab:sentiment_data}
  \begin{tabular}{ccccc}
    \toprule
    Category & All & Neutral & Positive & Negative \\
    \midrule
    Chinese & 17043 & 9450 & 4115 & 3479 \\
    English & 74163 & 60739 & 1093 & 12331 \\
    \bottomrule
  \end{tabular}
\end{table}

\begin{table*}[!t]
\centering
  \caption{Text with a different sentiment label}
  \label{tab:sentiment_display}
\setlength{\tabcolsep}{0.07cm}
  \begin{tabular}{p{12.3cm}cc}
    \toprule
    Text & Label & Probability \\
    \midrule
    Issued by London Marketing Academy awarded to REBECCA WEBB for the successful completion of the Social Media and Facebook Ads Course and Certification Exam. & Positive & 0.846 \\
    I bought doggy coins. I lost a lot of money, can you help me to black the project side account. & Negative & 0.989 \\
    ETHMiner is Ethereum miner with fully automatic process. Start earning Ethereum now! & Neutral & 0.788 \\
    \bottomrule
  \end{tabular}
\end{table*}

\subsubsection{Chinese text analysis}
\begin{figure}[!t]
  \centering
  \includegraphics[width=3in]{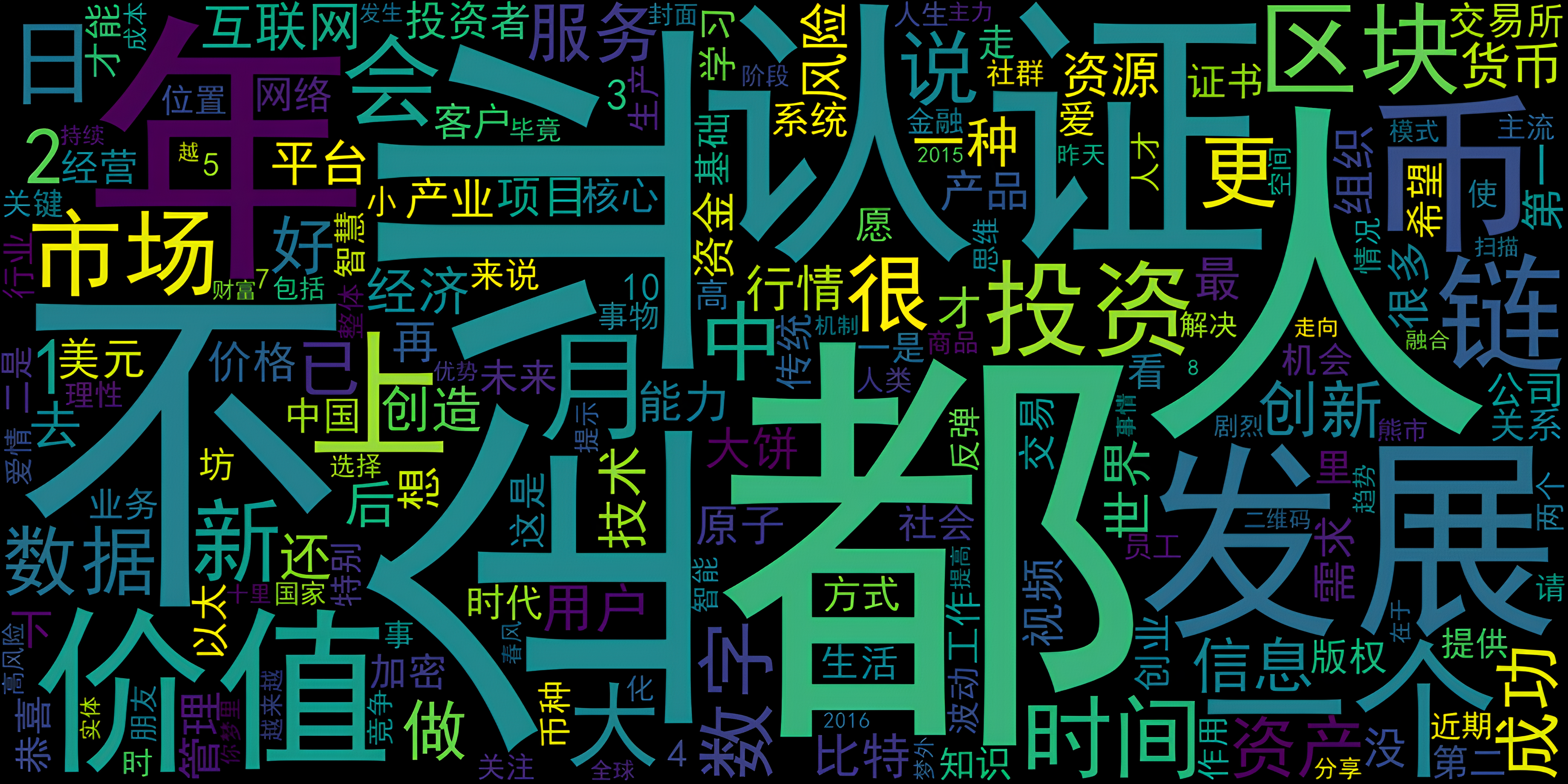}
  \caption{Word cloud visualization of Chinese textual data.}
  \label{ChineseWordCloud}
\end{figure}

After decoding, 17,043 eligible Chinese text files (occupying 14M space) were obtained. To understand the content of these text files from a macroscopic view, we cleaned and separated these texts into words and counted the word frequency. The word cloud is shown in Figure \ref{ChineseWordCloud}. Finally, we found that there are more words related to blockchain and finance, such as "coin", "development" and "investment". This implies that many text files are related to finance.

We further analyzed the text messages received by the active receiving addresses. Some addresses were found to receive only specific content, with the top five addresses that receive the most content mainly receiving information related to wishing, copyright certification, logistics traceability, financial analysis, and sexual harassment records. The most frequent content is related to wishing, with about 2,705 items in total being found. Similar content like blessings, love-related, and New Year wishes were also found.

In addition, in the sentiment classification experiment, we found that the text content with neutral and positive sentiments includes wishes, logistics traceability, and so on, but also records of sexual harassment, negative views on current affairs politics, social issues, and concerns about the COVID-19 pandemic, etc. At the same time, there has been some separatist political rhetoric. Due to the non-tamperable, transparent, and open nature of the blockchain, these statements will be permanently recorded and disseminated once they are stored in the blockchain.

\subsubsection{English type text analysis}
\begin{figure}[!t]
  \centering
  \includegraphics[width=3in]{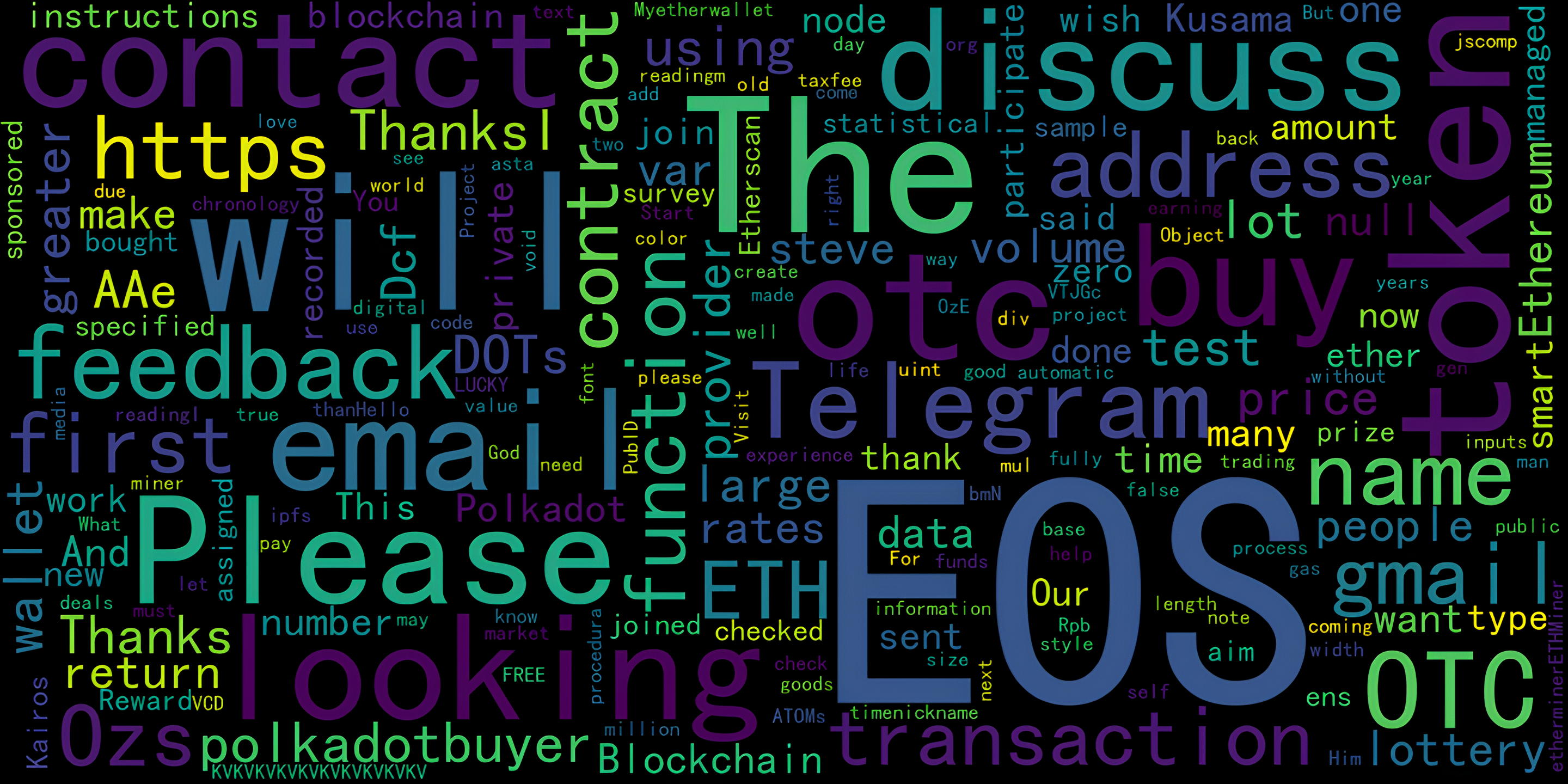}
  \caption{Word cloud visualization of English textual data.}
  \label{EnglishWordCloud}
\end{figure}

After decoding, 74163 (85.7M) English text files were obtained. The analysis of the English text files found that the most frequent content is about EOS tokens, with 31,958 records in total. 9,337 OTC-related records and 6,892 token offering/marketing advertisements were also found. For further analysis, the text content was cleaned, word separation was performed, and word frequency were counted. The word cloud is shown in Figure \ref{EnglishWordCloud}. The results showed that there were more words related to tokens, transactions, and smart contracts. "EOS" appeared more than 400,000 times, and "token" appeared more than 360,000 times, which mostly came from the crowdfunding process of token issuance. Of note, "OTC" appeared more than 240,000 times. This is probably because a large number of token offerings has led to a surge in transaction volume, which has led to a continuous rise in transaction fees and has also caused congestion in the Ethereum network, with more and more users choosing OTC transactions to avoid high fees.

In the sentiment classification of English text, texts with neutral and positive sentiments include token advertisements, token applications, blessings, and some event records, etc. It is worth noting that there are some racially discriminatory remarks in the negativity. At the same time, many users store the information of being cheated on the blockchain network to the chain, recording the incident and exposing the scam, for example, fake token investment scam information based on the Alpha Finance protocol, whose hash value of this transaction is 0x4ef4$\footnote{ 0x4ef4ba3f39507b784892df99735f627d6c9e66173c5b6c9e7a8}$. Alpha Finance is a cross-chain Defi protocol designed to connect users to various financial services. The address 0xc88f$\footnote{0xc88fdbcaa45142c0cdd1da4eea79ed40022f15da}$ purchased ALPHA tokens with ether invested by other users, but did not transfer the investment return to the corresponding users. This also verifies that there are frauds in the Ethereum network. This corroborates with the research on phishing scams~\cite{14} and Ponzi scheme identification~\cite{15} in Ethereum.

\vspace{0.2cm}
\framebox{%
\begin{minipage}{0.9\linewidth}
\textbf{Finding 4: }A large amount of text data is embedded in the Ethereum network, mainly including common content such as wishing, copyright certification, logistics traceability, financial analysis, etc., which shows that the use of blockchain for "evidence" has been recognized by people. In addition, the text also contains deceptive records, sexual assault records, and other information.
\end{minipage}}

\subsection{Sensitive information embedding analysis}
When analyzing the embedded text, we found a sensitive information embedding behavior that deserves attention: in October 2020, someone with ulterior motives wrote the personal information of several deputy ministerial-level cadres and members of the national leadership team into the Ethereum network, including ID number, phone number, email address, work and home addresses, etc. In order to understand the authenticity of this information, we verified the names and work units of most people through public channels, and the results show that the information is true. Although the ID number and mobile phone number cannot be further verified, it is very likely to be real as other information has been confirmed. Obviously, the leakage of such information has brought great dangers to the personal privacy and personal safety of leading state cadres. For privacy reasons, no reference is made here.

To gain a deeper understanding of the embedding behavior of such sensitive information, a more comprehensive analysis of the addresses where this sensitive information is embedded is performed. The address A (Address A for short. The first 4 digits of the address are 0x8562. To avoid further leakage of privacy, the full address is not provided here) had 15 transactions for embedding information in October 2020. The information is related to 29 officials in 12 departments, including the Ministry of Commerce of China, and the United Front Work Department of the CPC.

To further trace the source of funds, all transaction records from 10 million to 13 million blocks were first decoded, and then the transaction records of the three transfer addresses connected to the message uploading address were traced. Finally, the address B (0x7A34), which first transferred Ether to address A, was found to have suspicious behavior. Analysis of address B's transaction activities revealed that it was active between August 2020 and January 2021. The number of Ether balances peaked on November 27, 2020 (254.7 Ether), and then on December 23, 2020, all Ether from the account was transferred to other accounts in a scattered manner.

To discover more information, we further explored the B address-related transactions. First, the B address is sorted according to the amount of ether transferred out, and it is found that 7 transactions transferred more than 300Ether, and 55 transactions transferred more than 200Ether. We believe transactions of this account are suspicious. To find out the information related to the address, we searched the address B in Google and found it belongs to the ChangeNow exchange, and it was recorded on the Pastebin website $\footnote{https://pastebin.com/r7rJH0g2}$that the address participated in fund transfers after the KuCoin cryptocurrency exchange was hacked on September 25, 2020. Based on all these information, we speculate that the behaviors of address A in embedding sensitive information on the Ethereum network may have been funded by the hacker group behind it.
Embedding similar sensitive information on the Ethereum network poses challenges for individual privacy protection. If this behavior represents information transfer by lawbreakers, the potential dangers will be immeasurable, and it will also pose a threat to national security.

\vspace{0.2cm}
\framebox{%
\begin{minipage}{0.90\linewidth}
\textbf{Finding 5: }A malicious user was found to have embedded sensitive information of several important members of the leadership of relevant Chinese departments in Ether, and further analysis suggested that the source of funding involved hacking. While it is difficult to understand the motivation behind this act, it is a clear violation of the privacy of the people involved and may even threaten national security.
\end{minipage}}

\subsection{Information embedding network construction and analysis}
\subsubsection{Information embedding network construction}
To analyze the information embedding behavior of Ethereum from a macroscopic perspective, the Ethereum Information Embedding Network is constructed. First, we defined a set\(<from, to, input>\), consisting of information embedding transaction fields. Then we built an information embedding network (Information embed Network,\(IEN\)) based on this set. The detailed definitions are as follows:
\[ IEN = {(V,E,W)} | E = (v_{i},v_{j}),v_{i},v_{j}\epsilon V \]
Among this definition, \(V\) is the set of addresses of all embedded messages. \(E\) is a set of ordered edges formed by groups of information embedding behaviors between addresses. Each edge represents that the node \(v_{i}\) embeds information in the node \(v_{j}\). \begin{math}w:E\to \mathbb{R}^{+}\end{math}associate each edge with a weight that represents the sum of the number of embeddings. Therefore, \(IEN\) is a weighted directed graph.

\begin{figure*}[ht]
  \centering

  \subfloat[File data embedding network]{
    \includegraphics[width=0.22\textwidth]{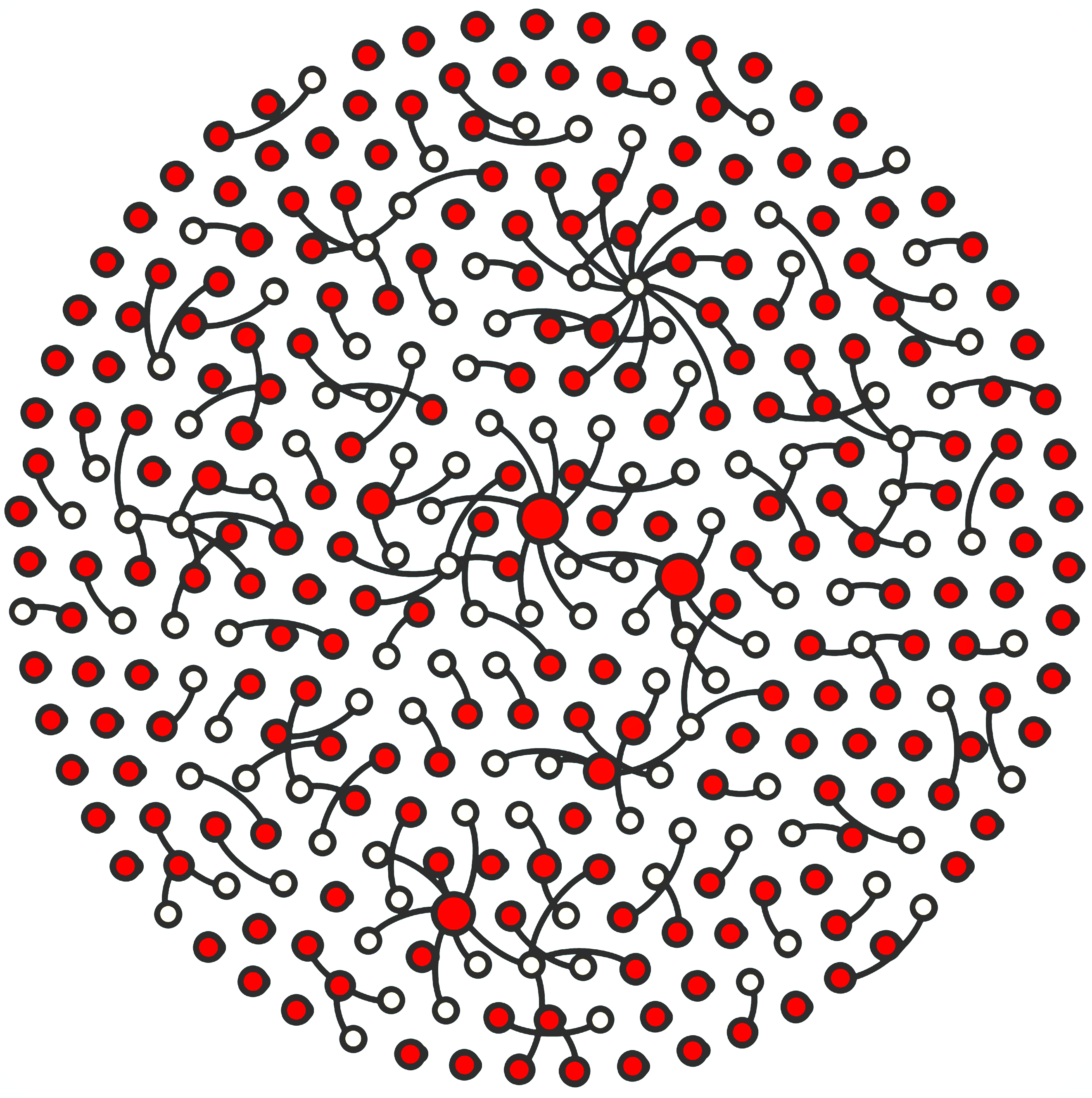}
    \label{fig9:a}
  }
  \hfill 
  \subfloat[Link data embedding network]{
    \includegraphics[width=0.22\textwidth]{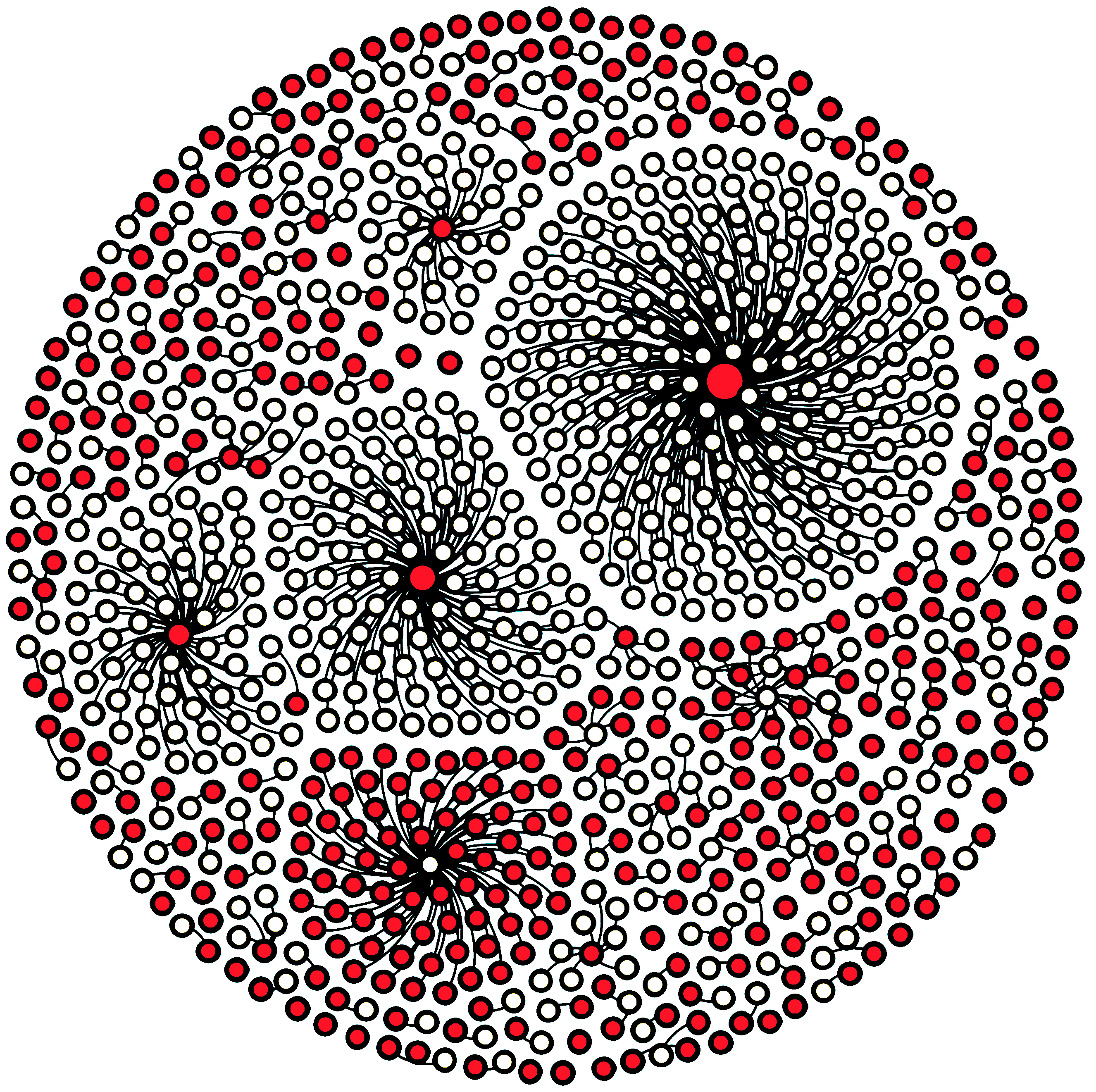}
    \label{fig9:b}
  }
  \hfill
  \subfloat[Chinese data embedding network]{
    \includegraphics[width=0.22\textwidth]{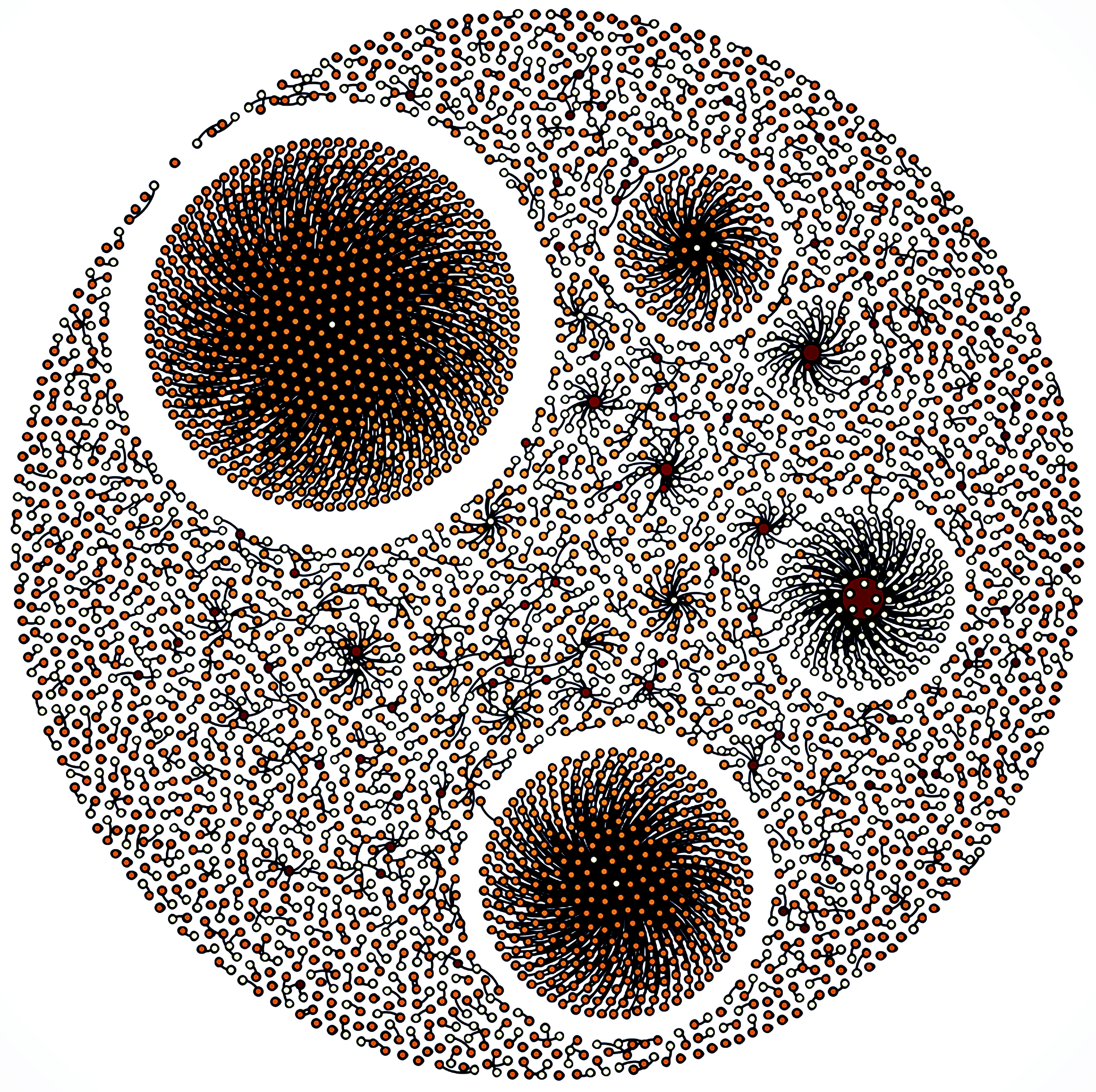}
    \label{fig9:c}
  }
  \hfill
  \subfloat[English data embedding network]{
    \includegraphics[width=0.22\textwidth]{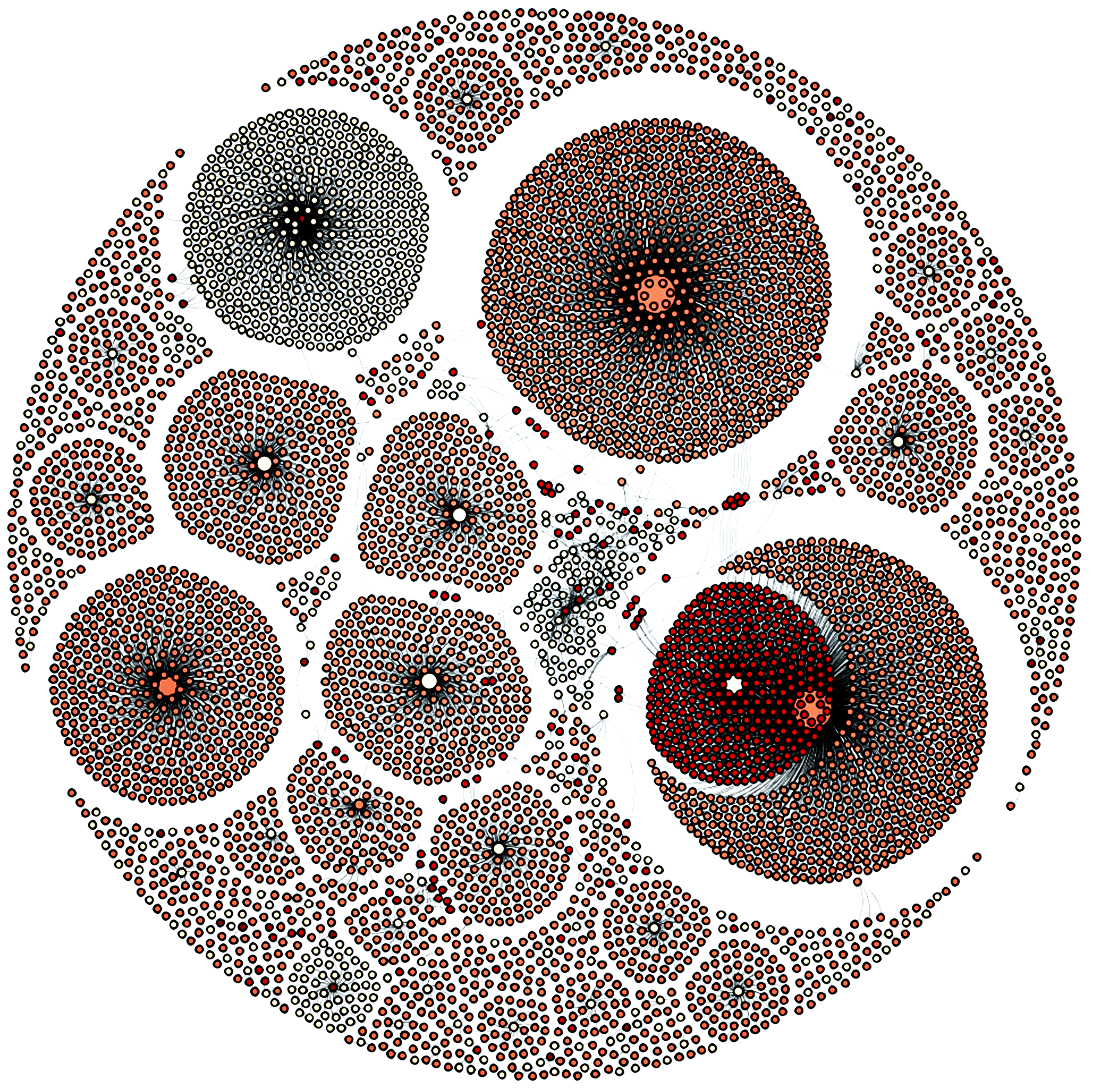}
    \label{fig9:d}
  }

  \caption{Information embedding network visualization: (a) File data embedding network, (b) Link data embedding network, (c) Chinese data embedding network, (d) English data embedding network.}
  \label{fig9}
\end{figure*}

Fig.~\ref{fig9} illustrates the information embedding network. Specifically, Fig.~\ref{a} depicts the file embedding network, Fig.~\ref{c} the Chinese embedding network, and Fig.~\ref{d} the English embedding network. Node size and color in the graphs are determined by the node degree distribution in the overall data embedding network, positively correlated with in-degree.

\subsubsection{\(IEN\) visualization result analysis}
The information embedding network contains a total of 44887 nodes and 91206 edges, which are divided into 4 different graphs according to the category of embedded information (Fig.\ref{fig9}). The analysis found that a part of the senders of the embedded information share the same address with the recipients, while another part of the addresses launched transactions with information embedding in the Ethereum network with other account addresses. In addition, some specific account addresses receive or send the same type of data or execute a certain task.

In the file embedding network diagram (shown in Fig.\ref{a}), most senders and receivers have the same address. When the recipient is the CreepyCryptos contract address, 30 addresses send NFT artwork to the contract. In the embedded network of link information (shown in Fig.\ref{b}), some users spread advertisements on a large scale through the Ethereum network, such as the podcast website with the URL "podcrypt. app" and the website with the URL "ddblock.io" Slideshow Software Inc.

In the information embedding network in Chinese (shown in Fig.\ref{c}) and English (shown in Fig.\ref{d}), a large number of information embedding addresses have an obvious clustering phenomenon. The analysis found that the first five address clustering phenomena embedded in Chinese information were used for copyright authentication, order records, logistics traceability, wishing blessings, and financial analysis; while in English texts, a large amount of embedded information was related to the issuance of tokens. Among them, the information related to EOS tokens constituted the highest proportion, with the number of detected items exceeding 30,000. Following it are FTC chain and ATOM token issuance advertisements and commodity transaction certification.

\begin{figure}[!t]
  \centering
  \includegraphics[width=3in]{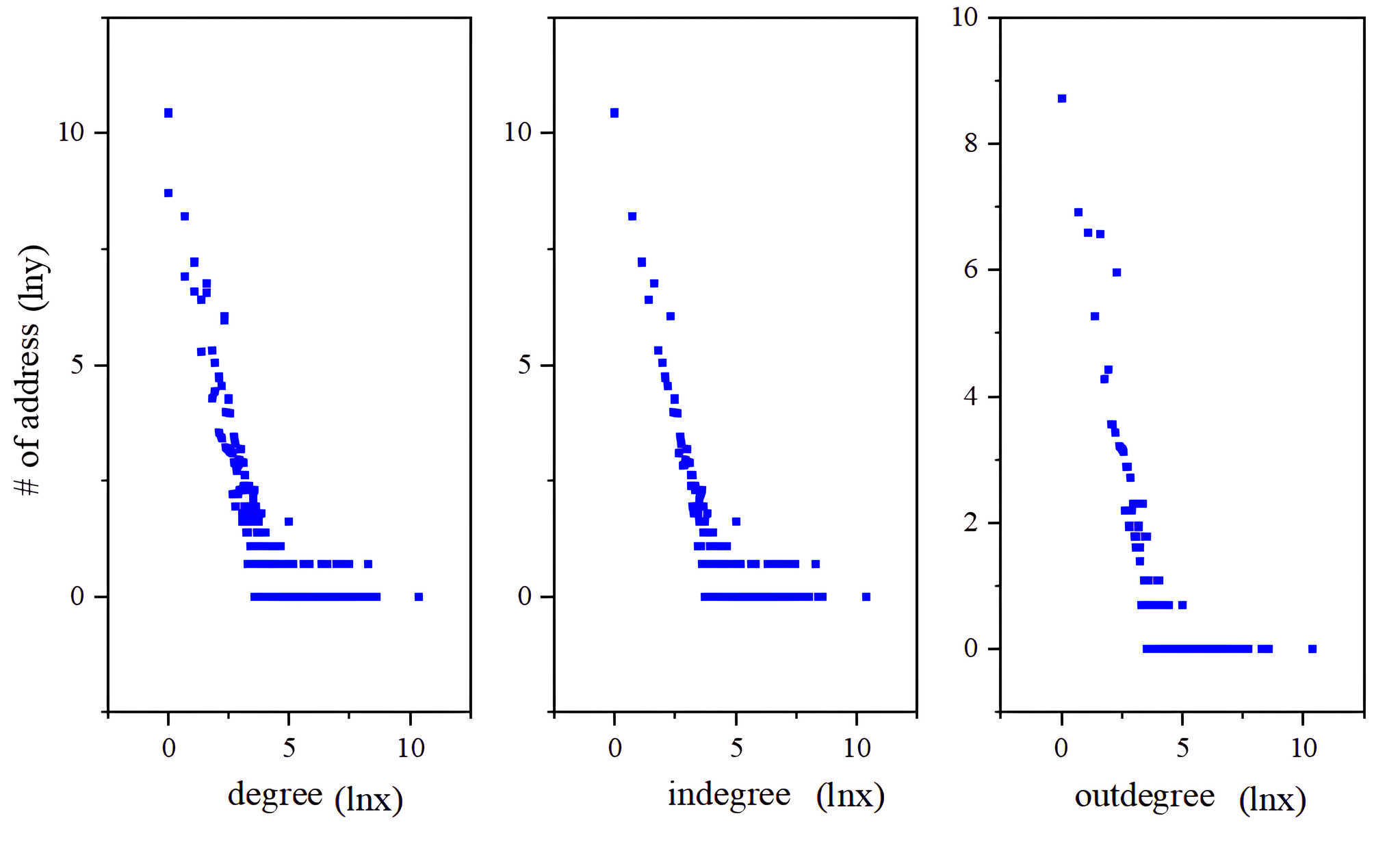}
  \caption{The degree, in-degree, and out-degree distribution of various types of information embedded in the Ethereum blockchain.}
  \label{powerlow}
\end{figure}

In addition, the total degree, the in-degree and out-degree distribution of various types of information embedded in the Ethereum blockchain were also analyzed. The degree distribution is shown in Figure \ref{powerlow}. Intuitively, all types of information-embedded graphs satisfy the power-law distribution. To save space, the relevant graphs are not shown here. A simple analysis found that 77\% of addresses embedded information in the chain only once, and 95\% of addresses embedded information less than 6 times. Thus, most addresses are inactive in embedding information in the Ethereum network, but there are still a small number of addresses that frequently embed information to receive or send the same type of data or perform a certain task. The address with the most in-degrees was 0xc68b$\footnote{0xc68b5d7d31993a8d53512da7bd4a63e64d096497}$, which received 31,958 texts and was used to receive information about EOS tokens. The address with the most out-degrees was 0x0920$\footnote{0x09205a61e79f9734ceff61a50e49c1dee362d623}$, which sent 4,407 text messages. The contents of this address were advertisements for two platforms, Polkadot (DOT) and Kusama (KSM). They are two independent platforms on which users can publish and operate their blockchains.

\vspace{0.2cm}
\framebox{%
\begin{minipage}{0.9\linewidth}
\textbf{Finding 6: }Of the addresses involved in message embedding in the Ethereum network, 77\% have embedded messages only once, and 95\% have embedded messages less than six times. However, there are cases where some addresses embed information frequently to receive or send the same type of data or perform a certain task.
\end{minipage}}

\section{Preventing harmful content in Ethereum: risks and measures analysis}
\label{6}
This article discusses the issue of embedded data in the Ethereum blockchain and aims to provide a detailed analysis of the topic. The study was able to restore 175 common file types, 296 image files, and 91,206 text files from approximately 300 million transactions and then analyzed the content. The research found that although most of the content on the Ethereum blockchain is harmless, there is also harmful content such as personal information, pornographic images, hate speech, and racial discrimination.

The data restoration algorithm and technology used in this study may offer new solutions to privacy and security issues on the blockchain. This research represents the first comprehensive exploration and analysis of embedded content in the Ethereum blockchain, providing important insights for the public to fully understand the technology. Furthermore, it may serve as a reference for blockchain regulatory agencies.In the following content, we will discuss the adverse effects of embedded data in the Ethereum network and propose several measures to reduce these risks.
\subsection{Potential risks analysis}
(1)Pornographic Images: The dissemination of pornographic images on the Ethereum network may negatively affect the healthy development of children and adolescents. Moreover, it may trigger moral controversies and even be exploited for malicious behaviors such as cyberbullying or sexual harassment.

(2)Privacy data: Personal privacy data embedded in the Ethereum network, such as social arrangements, health records, etc., may be used by hackers to use identity theft, fraud, and other malicious behaviors.

(3)Racial discrimination and divisive speech: Racially discriminatory language and behavior embedded in the Ethereum network can lead to injustice and inequality within the community, undermining its harmony and development, and even inciting conflicts and violent incidents. Similarly, divisive speech embedded in the Ethereum network, such as hate speech and racism, can exacerbate community conflicts and even lead to community division, thereby impacting the security and stability of the entire Ethereum network.

(4)Malware: Malware embedded in the Ethereum network can potentially cause network paralysis or severe damage to users' property security, as such software can spread widely on the network and potentially infect other nodes.

\subsection{ Measures to prevent embedding of harmful data in Ethereum}
(1)Pornographic Images: Automatic Detection and Filtering System based on Image Recognition Technology: By automatically identifying and filtering images uploaded to the Ethereum network, data containing pornographic images can be eliminated to prevent their spread. Moreover, it is possible to design a filtering protocol for Ethereum blockchain data embedding content~\cite{zhang2021approach}, aimed at preventing the inclusion of harmful content on the chain. This protocol could involve mechanisms for reviewing data either before or after embedding into the network, thereby deterring the embedding of malicious or irrelevant data. Additionally, an effective oversight and management framework must be established~\cite{gimenez2023cybersecurity,chen2022blockchain}.

(2)Privacy data: Cryptographic techniques can be used to encrypt private data, for example, using hash functions to encrypt user data to protect their privacy. Additionally, multi-party computation and other similar technologies can also be used in smart contracts to protect user privacy~\cite{chen2022blockchain}.

(3)Racial discrimination and divisive speech: The potential negative impact of embedding racial discrimination and divisive speech (such as hate speech and racism) in the Ethereum network can lead to community conflict and threaten the stability and security of the entire network~\cite{dwivedi2022metaverse}. To address this risk, one solution is to utilize natural language processing technology and content filtering systems to detect and remove inappropriate content. Another approach is to audit smart contracts related to discriminatory behavior, with the involvement of professional review organizations, to ensure fairness and security. Since smart contracts are crucial to the Ethereum network, eliminating any discriminatory practices is essential to maintain a safe and inclusive environment.

(4)Malware: To mitigate the embedding of malicious executable files in a blockchain, one effective approach is to establish an auditing mechanism that reviews uploaded files before they are added to the blockchain~\cite{gimenez2023malicious}. This mechanism can be implemented through a combination of manual and automated auditing. In addition, smart contracts can be utilized to restrict the types of files that can be uploaded, such as only allowing images and text files while disallowing executable files. Punitive measures can also be included in the smart contract to deter the uploading of malicious files.

In conclusion, the risks associated with embedding harmful data in Ethereum cannot be ignored, and it is essential to implement preventive measures to safeguard the platform's integrity and reputation. By taking a proactive approach to security, the Ethereum community can ensure the continued growth and success of the platform.

\section{Related work}
\label{3}
Since the birth of Bitcoin, there has been a large amount of literature based on blockchain data mining. Three studies are closely related to the research in this paper. 

The first category is to focus on mining the blockchain to reveal user characteristics in the system, from discussing user privacy issues~\cite{8,9}, to identifying various user behaviors~\cite{19,20}, and revealing illegal activities~\cite{21,23}.

The second category mainly discusses smart contracts, the core elements of blockchain 2.0. Since the deployment of smart contracts is also carried out through transactions, the study of smart contracts can be regarded as a special analysis of transaction-embedded information. Many related studies have been generated around smart contracts, such as contract deployment and defects~\cite{24}, code analysis~\cite{26,27}, and application security~\cite{30}. 

The third category, and most relevant to this study, is the mining of arbitrary data embedded in the blockchain. Reference~\cite{17}  studied the data embedded in the Bitcoin blockchain network based on the OP\_RETURN and Coinbase fields and found bad content that contained child pornographic images and violated the privacy of others. 

Reference~\cite{21}  proposes a countermeasure to prevent child pornography images from being embedded in the Bitcoin blockchain. Reference~\cite{33}  discusses the problem of “poisoning attacks” in the Ethereum blockchain and finds 154 files embedded in it. Three of the exe files have been proven to be malicious programs through experiments. Reference~\cite{32} studied the possibility of using the "value" field to construct an MHAC-based encryption algorithm for secret information embedding in Ethereum network transactions.

To the best of our knowledge, compared with previous work, this article presents a comprehensive analysis of the Ethereum data embedding problem for the first time. Specifically, this study focuses on data embedding on the Ethereum platform and proposes a novel data recognition and restoration algorithm designed specifically for Ethereum blockchain scenarios, which successfully restores a large amount of textual and image files. The study applies the FastText algorithm to classify the restored textual data based on sentiment, identifies and analyzes sensitive or illegal information embedded within it. Additionally, the study utilizes the open-source NSFWJS library to accurately identify inappropriate images with a 100\% accuracy rate. Finally, the potential risks associated with the discovered illicit content are analyzed, and corresponding regulatory measures are proposed. Overall, this study provides valuable insights into the public's comprehensive understanding of blockchain technology and can serve as a reference for blockchain regulatory agencies.

\section{ Conclusion and future work}
\label{7}
To fully explore the current status of data embedding in the Ethereum blockchain network, this paper explores the content of the "input" field in the Ethereum transaction record. About 3.4 billion transaction records were decoded after using the Ethereum client parity to get the raw data in-chain. To effectively restore data, a data restoration algorithm based on file feature code and text encoding was proposed to identify and restore the ordinary text, files, and images in blockchain transactions. Finally, 175 common types of files, 296 images, and 91206 text data were restored. Analyze the parsed information using text sentiment classification algorithms and image recognition methods. Through the analysis of various types of information , a variety of embedded sensitive and illegal information was found (please refer to each subsection for a summary of the findings), which points out the potential risks of the Ethereum network.

In future work, various measures can be implemented to tackle the challenges posed by the embedding of malicious data into blockchains. Firstly, the privacy protection mechanisms in blockchain technology can be further explored to identify ways to limit harmful data embedding while still safeguarding data privacy. Additionally, smarter data restoration algorithms can be developed to enhance accuracy and efficiency. More comprehensive review mechanisms, including real-time monitoring and detection using machine learning techniques, can be established to prevent the embedding of harmful data. Community education can be strengthened to increase awareness and prevent the embedding of malicious data into blockchains. Finally, a global collaborative mechanism can be established to jointly address the challenges posed by the embedding of malicious data into blockchains. This can include the formation of global regulatory standards and mechanisms.

\bibliographystyle{plainnat}
\bibliography{reference}

\vfill

\end{document}